\documentclass[3p,times,twocolumn]{elsarticle}

\usepackage{ecrc}

\volume{00}

\firstpage{1}

\journalname{Nuclear Physics B Proceedings Supplement}

\runauth{E. Fuchs}

\jid{nuphbp}

\jnltitlelogo{Nuclear Physics B Proceedings Supplement}

\usepackage{amssymb}

\biboptions{sort&compress}

\usepackage[figuresright]{rotating}

\usepackage{widetext}
\usepackage{lastpage}
\usepackage{subfigure}
\usepackage[colorlinks=true,citecolor=blue,urlcolor=blue,linkcolor=blue]{hyperref}

\newcommand{\co}{\tilde{\chi}_1^0}
\newcommand{\cf}{\tilde{\chi}_4^0}

\newcommand{\tb}{t_{\beta}}

\newcommand{\bpm}{\begin{pmatrix}}
\newcommand{\epm}{\end{pmatrix}}

\newcommand{\qsq}{q^{2}}
\newcommand{\psq}{p^{2}}

\newcommand{\Zbf}{\hat{\textbf{Z}}}
\newcommand{\Mm}{\mathcal{M}}
\newcommand{\Pp}{\mathcal{P}}
\newcommand{\Dd}{\mathcal{D}}

\newcommand{\BW}{\Delta^{\textrm{\scriptsize{BW}}}}
\newcommand{\BWs}{\Delta^{\textrm{\scriptsize{*BW}}}}
\newcommand{\Zb}{\hat{\textbf{Z}}}
\newcommand{\Zz}{\hat{Z}}

\begin{document}

\begin{frontmatter}



\dochead{\small{\hfill {\tt DESY 14-195}}}

\title{Interference effects of neutral MSSM Higgs bosons\\with a generalised narrow-width approximation}


\author{Elina Fuchs}
\address{DESY, Notkestr. 85, 22607 Hamburg, Germany\\
\rm{elina.fuchs@desy.de}}

\begin{abstract}
Mixing effects in the MSSM Higgs sector can give rise to a sizeable interference between the neutral Higgs bosons. On the other hand, factorising a more complicated process into production and decay parts by means of the narrow-width approximation (NWA) simplifies the calculation. The standard NWA, however, does not account for interference terms. Therefore, we introduce a generalisation of the NWA (gNWA) which allows for a consistent treatment of interference effects between nearly mass-degenerate particles. Furthermore, we apply the gNWA at the tree and 1-loop level to an example process where the neutral Higgs bosons $h$ and $H$ are produced in the decay of a heavy neutralino and subsequently decay into a fermion pair. The $h-H$ propagator mixing is found to agree well with the approximation of Breit-Wigner propagators times finite wave-function normalisation factors, both leading to a significant interference contribution. The factorisation of the interference term based on on-shell matrix elements 
reproduces the full interference result within a precision of better than $1\%$ for the considered process. The gNWA also enables the inclusion of contributions beyond the 1-loop order into the most precise prediction.
\end{abstract}


\end{frontmatter}


\section{Introduction}
\label{sect:intro}
\vspace{-0.06cm}
Many models of new physics predict additional particles beyond the standard model. Depending on the specific model and its underlying parameters, some of the new states might be nearly mass-degenerate, for example two of the neutral Higgs bosons or some sfermions of the MSSM. In such cases with a small mass gap between two intermediate particles and simultaneously a sizeable mixing among them, the interference term between contributions involving either of the nearly degenerate particles may become large. On the other hand, an extended particle spectrum can lead to long cascade decays of unstable particles. But many particles in the final state in conjunction with the need for precise loop corrections cause a technical challenge. So the well-known narrow-width approximation is useful to calculate the production and decay of an intermediate particle. This can be iterated until a complicated process is decomposed into sufficiently short sub-processes. Some Monte-Carlo generators make use of this procedure, and 
experimental searches for new particles are interpreted with respect to the prediction of a production cross section times branching ratio(s). Yet, the NWA in its standard version (sNWA) does not take interference terms into account. Instead of neglecting the interference or performing the full calculation, which is - especially at higher order - not always possible, we formulate a generalised NWA (gNWA) that also factorises the interference term on-shell and thereby enables the separate computation of loop corrections to the production and decay parts including the interference term\,\cite{Fuchs:2014ola}. Particularly, we apply the gNWA on interfering Higgs bosons within the decay of the neutralino $\cf$ into $\co$ and a $\tau$-pair. In a scenario with real MSSM parameters, only the two $\mathcal{CP}$-even Higgs bosons $h,H$ mix. But the gNWA can also be used for the $\mathcal{CP}$-violating interference between $H$ and $A$ in case of complex parameters and in the context of other new physics models.
\section{Generalised narrow-width approximation}\label{sect:gNWA}
\subsection{Standard NWA}
\begin{figure}[ht!]
\centering
 \includegraphics[width=0.8\columnwidth]{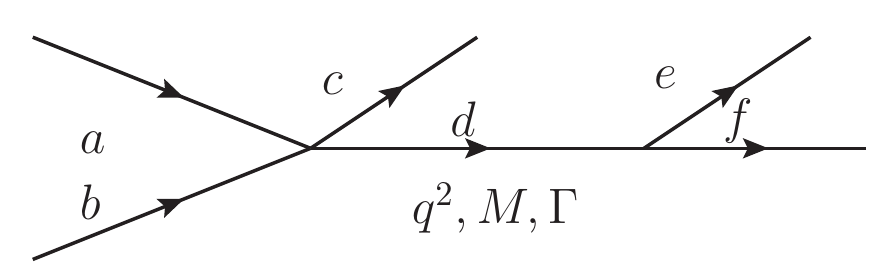}
\caption[Splitting a process into production and decay]{The resonant process $ab \rightarrow cef$ is split into the production $ab \rightarrow cd$ and decay $d\rightarrow ef$ with particle $d$ on-shell.}
\label{fig:prod_decay}
\end{figure}

The narrow-width approximation factorises a more complicated process into the on-shell production of an unstable particle and its subsequent decay. In Fig.\,\ref{fig:prod_decay}, the intermediate particle $d$ with mass $M$, total width $\Gamma$ and momentum $q^{2}$ is described by the Breit-Wigner propagator 
\begin{equation}
\BW\left(q^{2}\right):=\frac{1}{q^2 - M^2 + iM\Gamma} \label{eq:BWdef}~~.
\end{equation}
If its width is narrow, $\Gamma\ll M$, if the production and decay processes are kinematically open and if non-factorisable and interference contributions are negligible, the cross section of the generic example process $ab\rightarrow cef$ via the resonant state $d$ can be approximated by
\begin{equation}
 \sigma_{ab \rightarrow cef} \simeq \sigma_{ab \rightarrow cd} \times \textrm{BR}_{d\rightarrow ef}. \label{eq:NWAbasic}
\end{equation}

\subsection{Generalised NWA at tree level}
However, in the presence of several resonances, interference effects can be relevant if the contributing intermediate particles 1 and 2 are nearly degenerate, i.e., if their mass splitting is below one of the total widths, $|M_1-M_2|\leq \Gamma_1,\Gamma_2$. Denoting the production and decay matrix elements by $\Pp(q^{2})$ and $\Dd(q^{2})$, respectively, the full interference term in Eq.\,(\ref{eq:intexact}) can be approximated by on-shell matrix elements in the master formula in Eq.\,(\ref{eq:masterformula}),
\begin{widetext}
 \begin{eqnarray}
 \sigma_{\rm{int}} &=& \int \frac{d\Phi_P(q^{2}) dq^{2} d\Phi_D(q^{2})}{2\pi F}2\textrm{Re}\left\lbrace \Delta^{\textrm{\scriptsize{BW}}}_1(q^{2})\Delta^{*\textrm{\scriptsize{BW}}}_2(q^{2}) \Pp_{1}(\qsq)\Pp^*_{2}(\qsq)\Dd_{1}(\qsq)\Dd^*_{2}(\qsq)\right\rbrace\label{eq:intexact}\\
 &\simeq& \frac{2}{F}\textrm{Re}\left\lbrace\int \frac{dq^{2}}{2\pi}\BW_1(q^{2})\BWs_2(q^{2}) \left[\int d\Phi_P(q^{2})\Pp_{1}(M_1^{2})\Pp^*_{2}(M_2^{2})\right]\left[\int d\Phi_D(q^{2}) \Dd_{1}(M_1^{2})\Dd^*_{2}(M_2^{2}) \right]\right\rbrace\label{eq:masterformula},
\end{eqnarray}
\end{widetext}
where $\Phi_{P/D}$ are the production/decay phase spaces and $F$ the flux factor. Moreover, as a technical simplification beyond the on-shell approximation for matrix elements, also the phase spaces in Eq.\,(\ref{eq:masterformula}) can be evaluated on-shell and thus taken out of the $\qsq$-integral. Under the additional assumption of equal masses $M_1=M_2$, one can express the interference term through weight factors $R$, either as the weighted sum in Eq.\,(\ref{eq:intR}) or in terms of only one of the resonant particles in Eq.\,(\ref{eq:intRtilde}),  
\begin{eqnarray}
 \sigma_{int} &\simeq& \sigma_{P_i}\textrm{BR}_i\cdot R_i + \sigma_{P_j}\textrm{BR}_j\cdot R_j\label{eq:intR}\\
&\simeq& \sigma_{P_i}\, \textrm{BR}_i\cdot \tilde{R}_i, \hspace{1.7cm} i,j\in{1,2}\label{eq:intRtilde},\\
R_i &:=& 2M_i \Gamma_i w_i\cdot 2\textrm{Re}\left\lbrace x_i I \right\rbrace \label{eq:R},\\
\tilde{R}_i &:=& 2M_i \Gamma_i \cdot \textrm{Re}\left\lbrace x_i I \right\rbrace
 \label{eq:Rtilde}.
\end{eqnarray}
The $R$-factors are based on scaling factors $x$ as the ratio of couplings $C_{P/D}$ at the production/decay vertex\,\cite{Fowler:2010eba,ElinaMSc,Barducci:2013zaa}, the relative weight $w_i$ of each resonance $i$ and the overlap integral $I$ of the Breit-Wigner propagators within the kinematic boundaries $\qsq_{\rm{min}}, \qsq_{\rm{max}}$,
\begin{eqnarray}
 x_i &:=& \frac{C_{P_i}C_{P_j}^{*}C_{D_i}C_{D_j}^{*}}{|C_{P_i}|^{2}|C_{D_i}|^{2}}\label{eq:xi},\\
 w_i &:=& \frac{\sigma_{P_i}\,\textrm{BR}_i}{\sigma_{P_1}\,\textrm{BR}_1+\sigma_{P_2}\,\textrm{BR}_2}\label{eq:wi},\\
 I&:=&\int\limits_{q^{2}_{\rm{min}}}^{q^{2}_{\rm{max}}} \frac{dq^{2}}{2\pi}\BW_1(q^2)\,\BWs_2(q^2)\label{eq:defI}.
\end{eqnarray}
Hence, Eqs.\,(\ref{eq:masterformula},\ref{eq:intR},\ref{eq:intRtilde}) add a new term to the standard NWA, resulting in the generalised NWA which includes the possibility of several resonances as well as their interference based on on-shell matrix elements or weight factors at the tree level.

\subsection{Generalised NWA at higher order}
In addition, the gNWA can be extended to incorporate higher-order corrections as long as non-factorisable contributions between the initial and final state such as box diagrams are negligible. At 1-loop order with virtual corrections to the production and virtual and real corrections to the decay, the product of matrix elements is expanded as
\begin{eqnarray}
 \Pp_1\Pp_2^{*} &\longmapsto& \Pp_1^{1}\Pp_2^{0*}+\Pp_1^{0}\Pp_2^{1*},\\
\Dd_1\Dd_2^{*} &\longmapsto& \Dd_1^{1}\Dd_2^{0*}+\Dd_1^{0}\Dd_2^{1*}+\delta_{\textrm{\scriptsize{SB}}}\Dd_1^{0}\Dd_2^{0*},
\end{eqnarray}
where the superscripts $0/1$ denote the tree/1-loop level on-shell matrix elements and $\delta_{\scriptsize{\textrm{SB}}}$ the factor of soft bremsstrahlung (see e.g. Ref.\,\cite{Denner:1991kt,FormCalc}).

The result stays UV- and IR-finite, if the diagrams containing virtual photons and $\delta_{\scriptsize{\textrm{SB}}}$ are evaluated at the same mass \cite{Fuchs:2014ola,Denner:1997ia,Grunewald:2000ju,Denner:2000bj}. It is possible to formulate the $R$-factor approximation at 1-loop order such that higher-order cross sections and branching ratios, but only tree level couplings are used\,\cite{Fuchs:2014ola}. On top of 1-loop diagrams, also corrections of higher order can be included into the gNWA,
\begin{equation}
\hspace{-0.5cm}
 \sigma^{\textrm{\scriptsize{best}}} =\sigma_{\textrm{\scriptsize{full}}}^{0}+\sum_{i}\left(
\sigma_{P_i}^{\textrm{\scriptsize{best}}}\textrm{BR}_i^{\textrm{\scriptsize{best}}}-\sigma_{P_i}^{0}\textrm{BR}_i^{0}\right)+
 \sigma^{\textrm{\scriptsize{int}}1+}\label{eq:Mbest},
\end{equation}
where the tree level cross section $\sigma^{0}_{\textrm{\scriptsize{full}}}$ avoids an uncertainty from factorisation at lowest order. The \textit{best} production cross section $\sigma_{P_i}^{\textrm{\scriptsize{best}}}$ and branching ratios $\textrm{BR}_i^{\textrm{\scriptsize{best}}}$ mean the sum of the tree level, strict 1-loop and all available higher-order contribution to the respective quantity. Therefore, the products of tree level production cross sections and branching ratios are subtracted because their unfactorised counterparts are already contained in the full tree level term $\sigma_{\rm{full}}^{0}$. The corrections to the interference term at the 1-loop level and beyond are denoted by $\sigma^{\textrm{\scriptsize{int}}1+}$.
\section{Mixing effects in the Higgs sector}
For an application of the gNWA to interference effects in the Higgs sector, we briefly review, based on Refs.\,\cite{Frank:2006yh,Williams:2011bu}, the mixing of MSSM Higgs bosons. Self-energy contributions to the mass matrix
\begin{equation}
 \textbf{M}_{ij}(\psq) = \delta_{ij}\,m_i^{2} - \hat{\Sigma}_{ij}(\psq)\label{eq:Mij},
\end{equation}
where $m_i$ denotes the tree level mass of $i,j=h,H,A$, are related to mixing between the neutral Higgs boson propagators $\Delta_{ij}(\psq)$. In the case of mixed external Higgs bosons, finite normalisation factors are introduced for correct on-shell properties and the proper normalisation of the S-matrix in the $\overline{\rm{DR}}$- or other renormalisation schemes without on-shell renormalisation conditions:
\begin{equation}
 \hat{Z}_{i} = \frac{1}{\frac{\partial}{\partial \psq}\frac{i}{\Delta_{ii}}}\bigg\vert_{\psq=M^{2}_{c_{h_a}}},
\hspace*{0.5cm}
\hat{Z}_{ij} = \frac{\Delta_{ij}(\psq)}{\Delta_{ii}(\psq)}\bigg|_{\psq=M^{2}_{c_{h_a}}}\label{eq:Zij},
\end{equation}
where $h_a,~a=1,2,3,$ are the mass eigenstates with loop-corrected masses $M_{h_a}$ and total widths $\Gamma_{h_a}$.
Close to the complex pole $M_{c_{h_a}}^{2} = M_{h_a}^{2} - i M_{h_a}\Gamma_{h_a}\label{eq:HComplexPole}$, 
the full momentum dependence of the internal propagators can be approximated by a combination of Breit-Wigner propagators and 
$\Zbf$-factors $\Zb_{ij}=\sqrt{\Zz_i}\Zz_{ij}$, evaluated at the complex pole\,\cite{Fowler:2010eba,HiggsMix:InPrep}.
Thus, we determine the interference contribution in terms of $\Zbf$-factors and Breit-Wigner propagators.
\section{Application of the gNWA at tree level}
\subsection{Example process: $\tilde{\chi}_4^{0}\rightarrow \tilde{\chi}_1^{0} \tau^{+}\tau^{-}$}
\begin{figure}[ht!]
\centering
\subfigure[3-body decay]{\includegraphics[width=0.9\columnwidth]{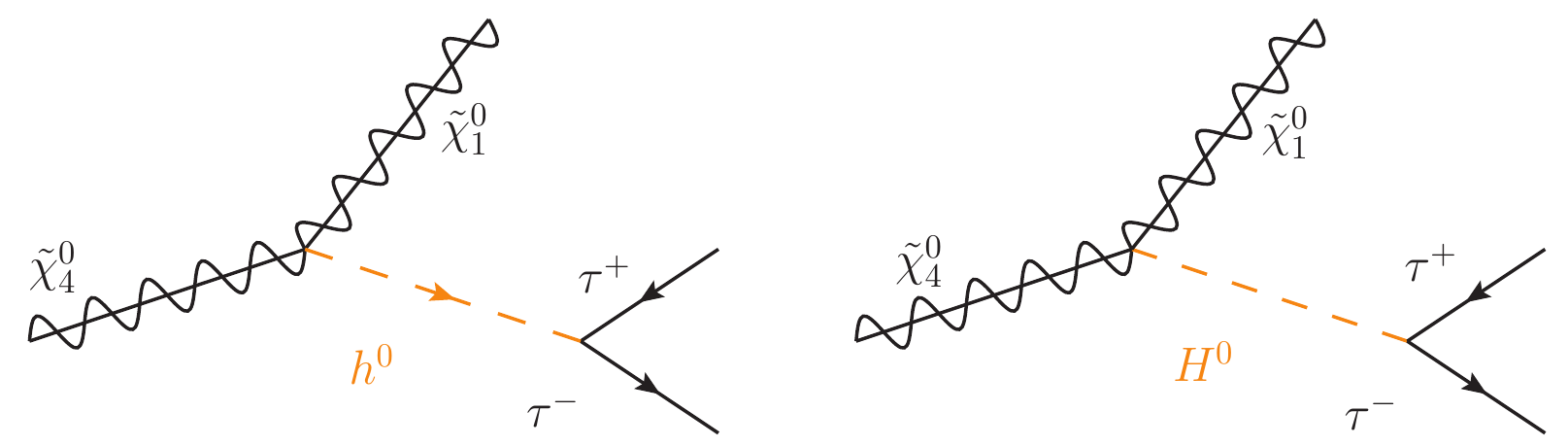}\label{fig:chi04_1to3}}\\
\subfigure[2-body decays]{\includegraphics[width=0.9\columnwidth]{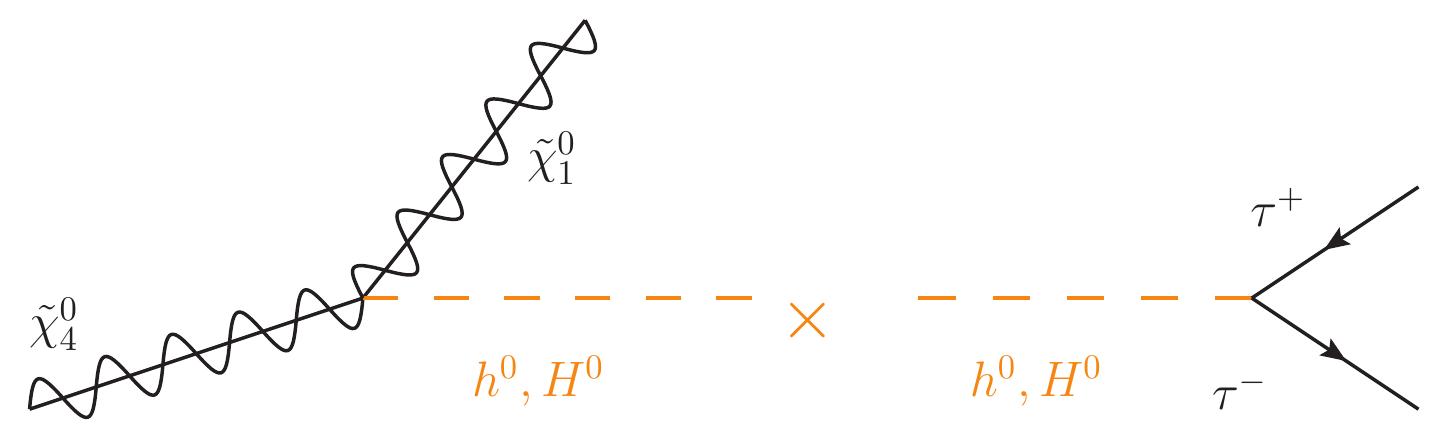}\label{fig:chi04_1to2}}
\caption[3-body decay of $\co$ split into 2-body decays]{Example process $\cf \rightarrow \co \tau^+ \tau^-$ with $h$ or $H$ as intermediate particles in the two interfering diagrams. The process is either considered as \textbf{(a)} a 3-body decay or \textbf{(b)} decomposed in two 2-body decays.}
\label{fig:chi04decay}
 \end{figure}

\begin{figure}[ht!]
 \begin{center}
  \subfigure[]{\includegraphics[width=0.65\columnwidth]{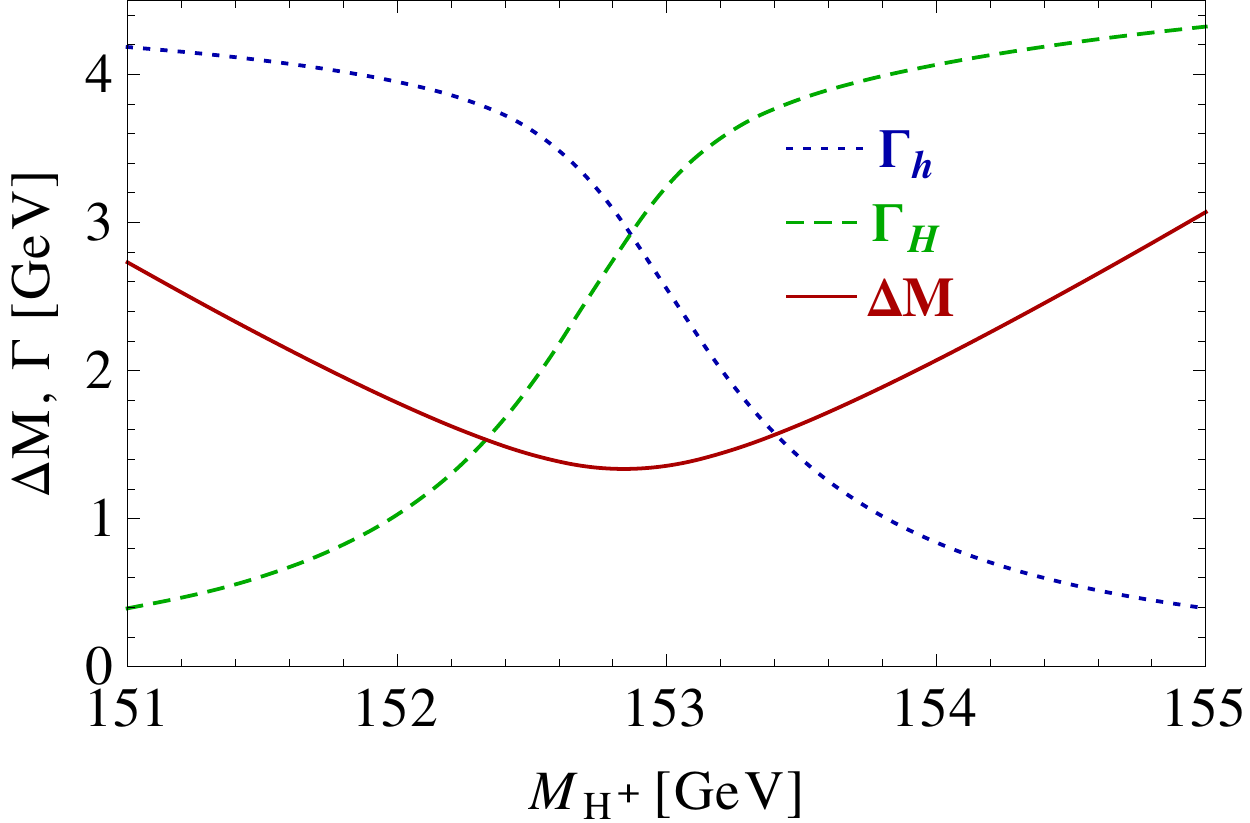}\label{fig:GM}\label{fig:DMGamma}}
  \subfigure[]{\includegraphics[width=0.67\columnwidth]{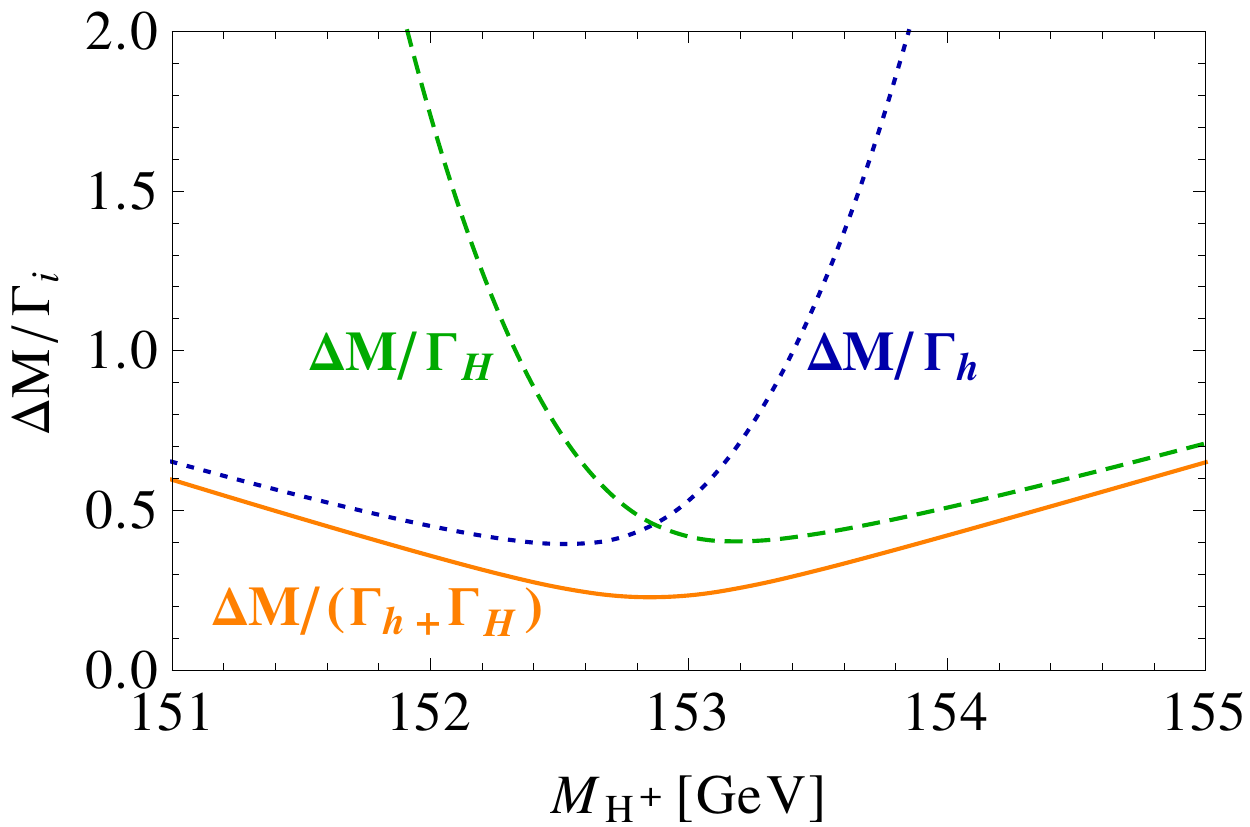}\label{fig:dmg}}
\caption{Higgs masses and widths from \texttt{FeynHiggs} including dominant 2-loop corrections in the modified $M_{h}^{\rm{max}}$ scenario. \textbf{(a):} Mass difference $\Delta M\equiv M_H-M_h$ (red) compared to total widths $\Gamma_h$ (blue, dotted) and $\Gamma_H$ (green, dashed). \textbf{(b):} Mass difference $\Delta M$ divided by total width of $h$ (blue, dotted), $H$ (green, dashed) and by the sum of both widths (orange).}
\label{fig:GammaDeltaM}
 \end{center}
\vspace*{-0.4cm}
\end{figure}

In order to investigate the possible impact of interference terms, we study the example process $\cf\rightarrow\co\tau^{+}\tau^{-}$ via a resonant $h$ or $H$. We confront the 3-body decay (Fig.\,\ref{fig:chi04_1to3}) with the prediction of the sNWA and gNWA based on 2-body production and decay parts (Fig.\,\ref{fig:chi04_1to2}) in a modified $M_h^{\rm{max}}$ scenario defined in Tab.\,\ref{tab:scenario}, where $h$ and $H$ are nearly degenerate. Their mass difference $\Delta M=M_H-M_h$, shown in red in Fig.\,\ref{fig:DMGamma}, is below one of the total widths $\Gamma_{h/H}$. The ratio $\Delta M/(\Gamma_h+\Gamma_H)$ in Fig.\,\ref{fig:dmg} gives a good indication of the parameter region of most significant interference. If it is minimal, the Breit-Wigner propagators overlap most strongly, causing a large overlap integral $I$.
\begin{table}[ht!]
 \begin{center}
  \begin{tabular}{|c c c c|}
   \hline
$M_1$& $M_2$& $M_3$& $M_{\textrm{\scriptsize{SUSY}}}$ \\
100\,GeV& 200\,GeV & 800\,GeV & 1\,TeV \\ \hline\hline
 $X_t$ & $\mu$&$\tb$& $M_{H^{\pm}}$\\
2.5\,TeV & 200\,GeV& 50& (153\,GeV) \\\hline
  \end{tabular}
\label{tab:scenario}
\caption[Parameters in the numerical analysis]{The modified $M_h^{\rm{max}}$ scenario used for the numerical analysis. Brackets indicate variation around this central value.}
 \end{center}
\end{table}

\subsection{Numerical analysis of the $h-H$-interference}
\begin{figure}[ht!]
 \begin{center}
  \includegraphics[width=\columnwidth]{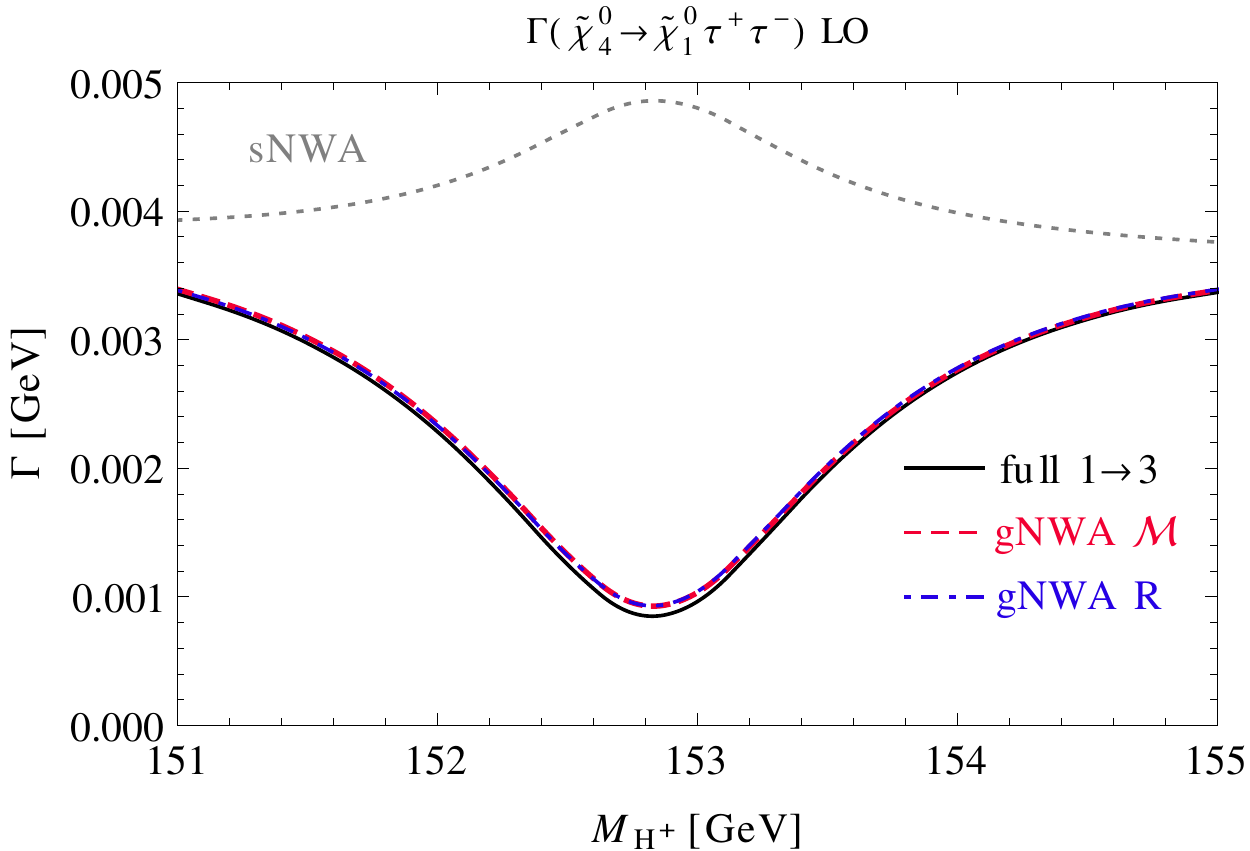}
\caption{The 1$\rightarrow$3 decay width of  $\cf \rightarrow \co \tau^{+}\tau^{-}$ at tree level with contributions from $h, H$ including their interference (black) confronted with the NWA: sNWA without the interference term (grey, dotted), gNWA including the interference term based on on-shell matrix elements denoted by $\Mm$ (red, dashed) and on the R-factor approximation denoted by R (blue, dash-dotted).}
\label{fig:sgNWA_tree}
 \end{center}
\vspace*{-0.5cm}
\end{figure}

As a function of the input Higgs mass $M_{H^{+}}$, Fig.\,\ref{fig:sgNWA_tree} 
shows the decay width of $\tilde{\chi}_4^{0}\rightarrow \tilde{\chi}_1^{0} \tau^{+}\tau^{-}$ computed with \texttt{FeynArts}\,\cite{Kublbeck:1990xc, Denner:1992vza, Kublbeck:1992mt, Hahn:2000kx, Hahn:2001rv} and \texttt{FormCalc}\,\cite{Hahn:1998yk, Hahn:1999wr, Hahn:2000jm, Hahn:2006qw, Hahn:2006zy} at the tree level, but improved by 2-loop Higgs masses, widths and $\hat{\textbf{Z}}$-factors from \texttt{FeynHiggs}\,\cite{Heinemeyer:1998np, Heinemeyer:1998yj, Degrassi:2002fi, Heinemeyer:2007aq}. 
The sNWA prediction (grey, dotted) overestimates the 3-body decay width (black) by up to a factor of 5 whereas the gNWA based on on-shell matrix elements (red, dashed, denoted by $\mathcal{M}$) or approximated by interference weight factors (blue, dash-dotted, denoted by $R$) reproduces the unfactorised result within a few percent uncertainty. 
The huge discrepancy between the sNWA and the 3-body decay width originates from the large, destructive interference term, owing to a small mass splitting $\Delta M$ and substantial mixing effects.

\section{Application of the gNWA at higher order}
\subsection{3-body neutralino decay at the 1-loop level}
\begin{figure}[ht!]
\begin{center}
\includegraphics[height=1.5cm]{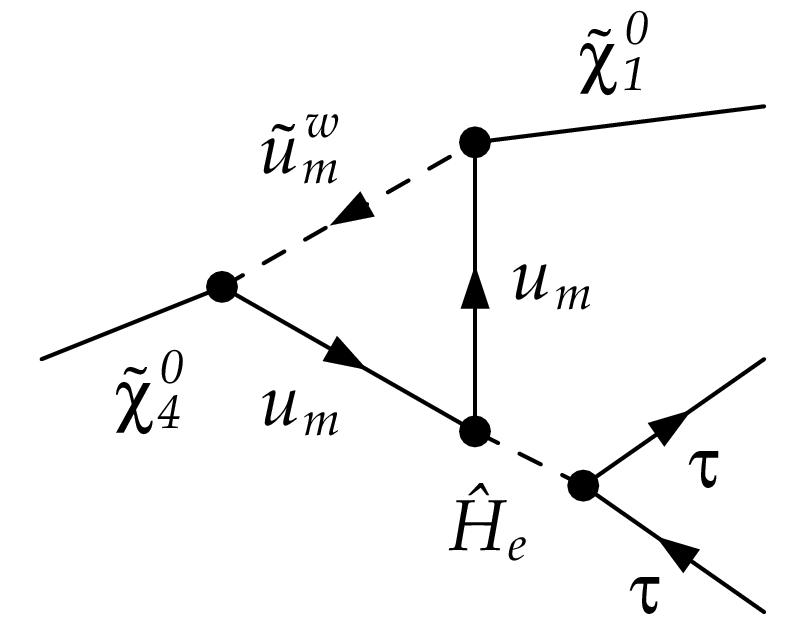}
\includegraphics[height=1.5cm]{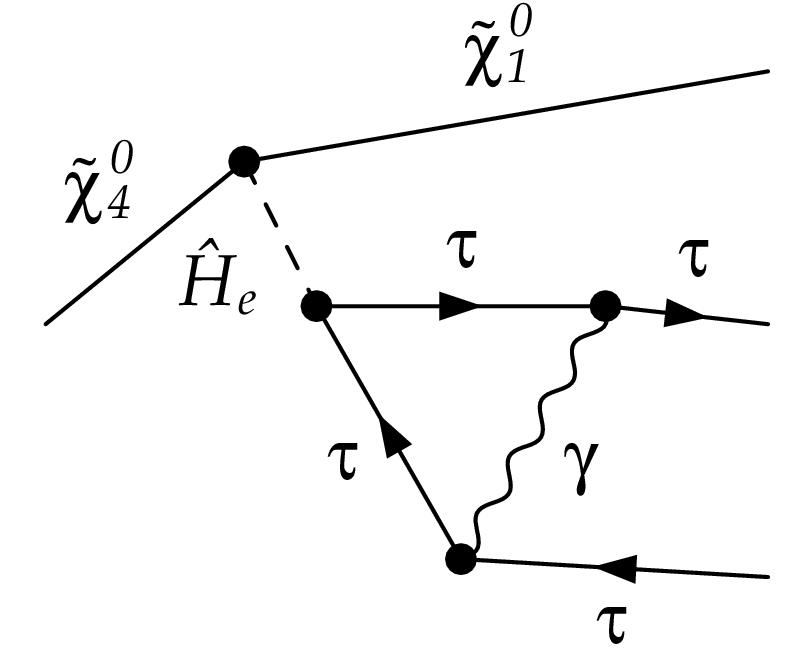}
\includegraphics[height=1.5cm]{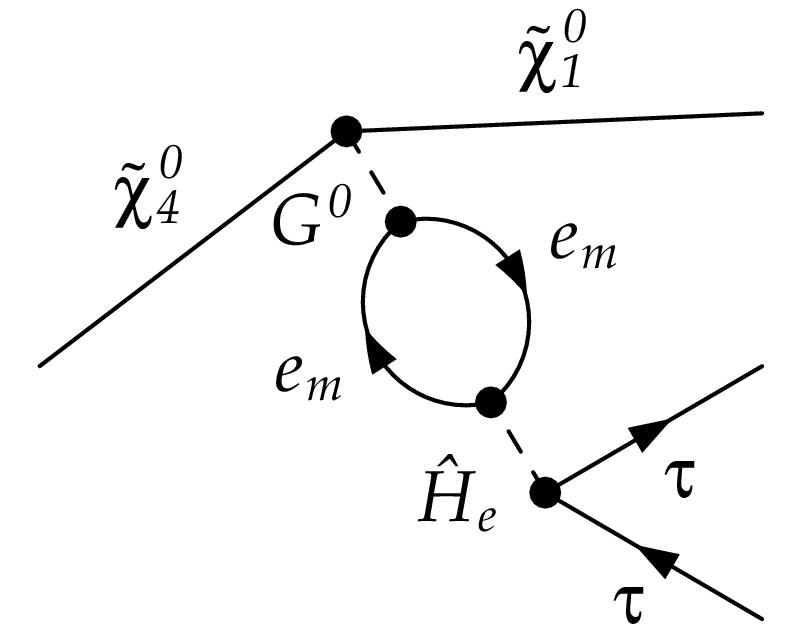}
\includegraphics[height=1.5cm]{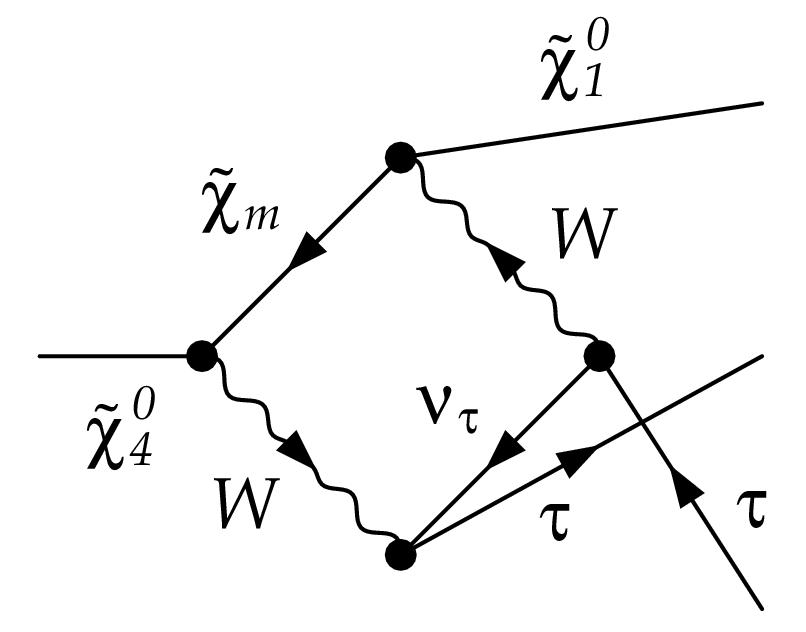} 
\caption{Example diagrams of the 3-body decay at 1-loop order: production and decay vertex, Goldstone-Higgs mixing, box.}
\label{fig:Loop13}
\end{center}
\vspace*{-0.5cm}
\end{figure}

In order test the gNWA at the 1-loop level, also the 3-body decay width is needed at this order as a reference. Vertex corrections at both vertices, Higgs self-energies as well as box diagrams and real photon emission contribute. The loop integrals are calculated with \texttt{LoopTools}\,\cite{Hahn:1998yk,Hahn:2010zi}. For illustration, some example diagrams are depicted in Fig.\,\ref{fig:Loop13}.

The neutralino-neutralino-Higgs vertex is renormalised in an on-shell scheme, see e.g. Refs.\,\cite{Lahanas:1993ib,Pierce:1994ew,Eberl:2001eu,Fritzsche:2002bi,Fowler:2009ay,Bharucha:2012nx,Bharucha:2012re}. Selecting the most bino-, higgsino- and wino-like states on-shell, in this scenario $\tilde\chi^{0}_{1,3,4}$, defines a stable renormalisation scheme\,\cite{Chatterjee:2011wc}. Consequently, the masses of the remaining electroweakinos $\tilde{\chi}^{0}_2$ and $\tilde{\chi}^{\pm}_{1,2}$ receive loop corrections. The Higgs sector is renormalised in a hybrid on-shell and $\overline{\rm{DR}}$-
scheme. While Higgs-Higgs mixing is already contained in the $\Zbf$-factors, the self-energies of Higgs and Goldstone/Z-bosons are calculated explicitly. Soft bremsstrahlung (SB) is proportional to the tree level width $\Gamma^{0}$, $\
\Gamma_{\textrm{\scriptsize{SB}}}=\delta_{\textrm{\scriptsize{SB}}}\,\Gamma^{0}$. IR-divergences arising from virtual photons in the final state vertex and from soft bremsstrahlung off the charged leptons in the final state cancel each other. The relative loop contribution $\Gamma_{\textrm{\tiny{full}}}^{1}/\Gamma_{\textrm{\tiny{full}}}^{0}-1$ is displayed in black in Fig.\,\ref{fig:gNWAMR1_prec_loopsize}, amounting to up to $11\%$.

\subsection{Validation of the gNWA including loop corrections}
\begin{figure}[ht!]
\vspace*{-0.2cm}
 \begin{center}
  \includegraphics[width=\columnwidth]{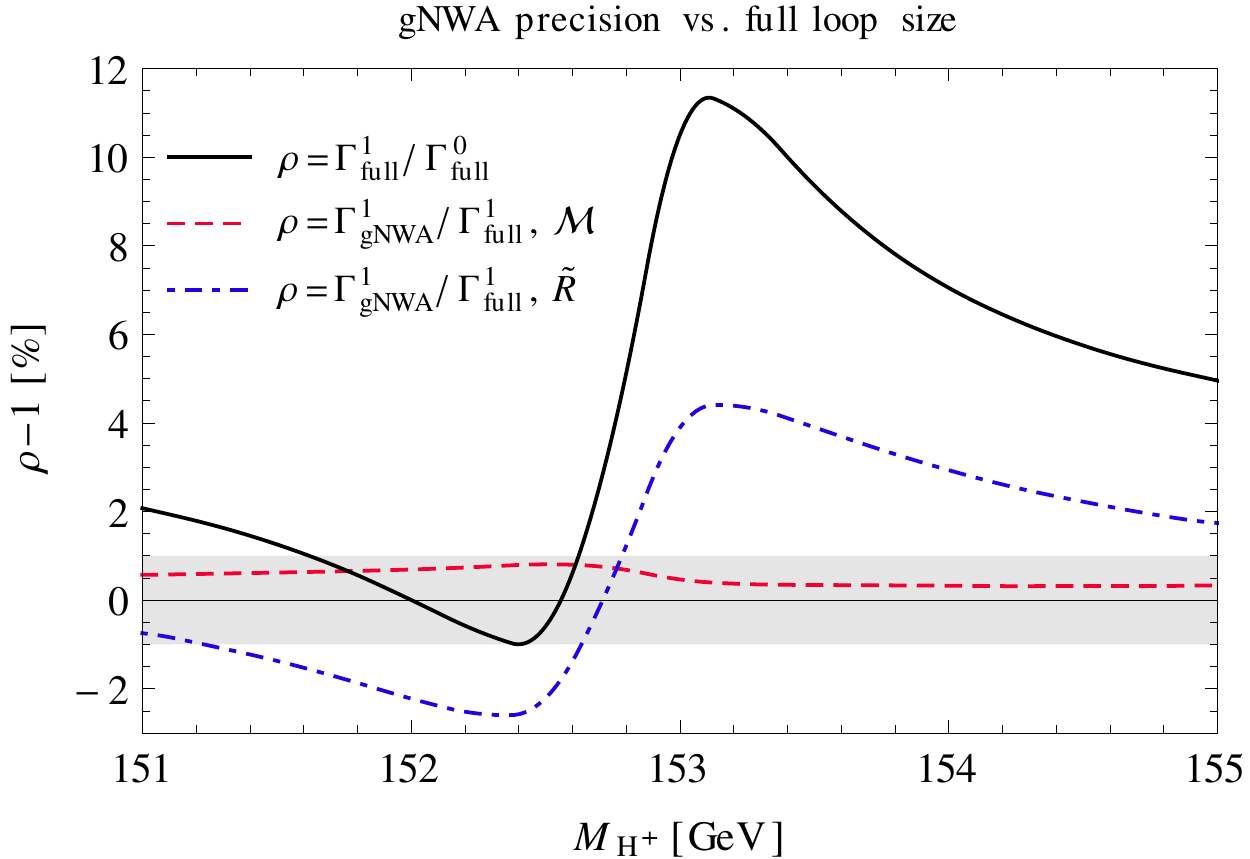}
\caption{Precision of the gNWA at the 1-loop level using the matrix element method denoted by $\Mm$ (red, dashed) and using the $R$-factor approximation denoted by $\tilde{R}$ (blue, dash-dotted) compared to the relative size of the loop contribution in the full calculation (black). The $\pm1\%$ region is indicated in grey.}
\label{fig:gNWAMR1_prec_loopsize}
 \end{center}
\vspace*{-0.5cm}
\end{figure}

\begin{figure}[ht!]
 \begin{center}
\includegraphics[width=8.2cm]{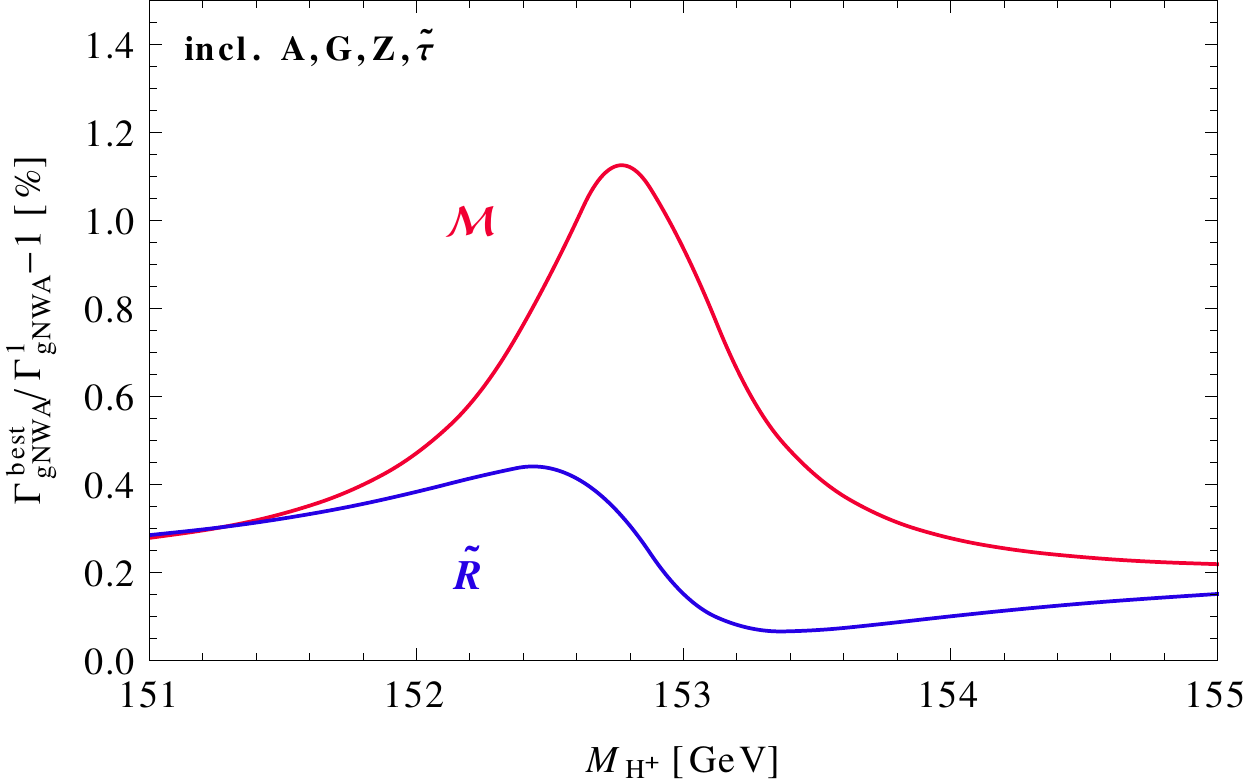}\label{fig:gNWAbestrel}
 \caption{The relative effect of the most precise branching ratios and product of 1-loop terms on the prediction of the gNWA with on-shell matrix elements (red, denoted by $\Mm$) and the R-factor approximation (blue, denoted by $\tilde{R}$) with respect to the 1-loop expansion.}
\label{fig:gNWAbest}
 \end{center}
\vspace*{-0.5cm}
\end{figure}

For the gNWA prediction, the 2-body decay widths $\Gamma(\cf\rightarrow\co h/H),~\Gamma(h/H\rightarrow\tau^{+}\tau^{-})$ as well as the production and decay on-shell matrix elements are calculated at the 1-loop level. As before, Higgs masses, total widths and $\Zbf$-factors are obtained from \texttt{FeynHiggs} at the leading 2-loop order. Fig.\,\ref{fig:gNWAMR1_prec_loopsize} shows the relative deviation $\Gamma_{\textrm{\scriptsize{gNWA}}}^{1}/\Gamma_{\textrm{\scriptsize{full}}}-1$ between the gNWA prediction with 1-loop corrections and the 1-loop 3-body decay width. While the $R$-factor approximation deviates from the full result by up to $4\%$, the method of on-shell matrix elements agrees with $\Gamma_{\textrm{\scriptsize{full}}}$ within a precision of better than $1\%$, which is of the order of the estimated remaining uncertainty of the full 1-loop result. Hence, the gNWA uncertainty of the $\mathcal{M}$-version stays mostly below the relative loop contribution, except where the full loop correction is 
minimal and even vanishing. The $R$-factor simplification can be regarded as a 
technically easier estimate of the interference term with loop corrections, whereas the $\mathcal{M}$-method provides a precise approximation of combined interference and higher order effects.
\\
The 3-body decays mediated by a resonant $\mathcal{CP}$-even Higgs boson $A$, a neutral Goldstone $G$, a $Z$-boson or a non-resonant $\tilde\tau$ have been omitted so far (but $H-G$ and $H-Z$ mixing has been included) since they do not interfere with the $\mathcal{CP}$-even Higgs bosons $h$ and $H$. For a realistic prediction of the example process, the lowest order contributions from $A,\,G,\,Z$ and $\tilde\tau$ are taken into account, increasing the width $\Gamma(\cf\rightarrow\co\tau^{+}\tau^{-})$ by an approximately constant shift of $4.15\cdot10^{-4}\,$GeV. Besides, loop contributions beyond the 1-loop level can be included in both versions of the gNWA according to Eq.\,(\ref{eq:Mbest}), such as branching ratios from \texttt{FeynHiggs} at the leading 2-loop level and  products of 1-loop partial results. Fig.\,\ref{fig:gNWAbest} presents the relative impact of these higher-corrections with respect to the 1-loop expansion of cross sections times branching fractions or matrix elements. In this example 
process and scenario, the impact amounts to up to $1.2\%$ for the $\mathcal{M}$-method and up to $0.4\%$ for the $R$-approximation.
\section{Conclusion}
Interference effects can drastically modify the cross section or partial decay width of a process if nearby resonances of particles with considerable mixing overlap within their total widths. Such degeneracies are possible in many BSM scenarios. We introduced a generalisation of the NWA to include the interference term in an on-shell approximation that maintains the convenient factorisation into a production cross section and branching ratio as in the standard NWA. For the example process of $\cf\rightarrow\co\tau^{+}\tau^{-}$ via $h/H$, the gNWA is validated against the 3-body decay width at lowest order and with 1-loop vertex, self-energy and box corrections and soft bremsstrahlung. In the analysed scenario of similar masses $M_h,\,M_H$ and large $h-H$ mixing, a significant destructive interference causes a huge discrepancy between the standard NWA and the full result. In contrast, the gNWA based on on-shell matrix elements in the interference term reproduces the full result within an accuracy of better 
than $1\%$. 
The factorisation of a more complicated process into a production and decay part achieved with the gNWA can be exploited for incorporating further higher-order contributions, leading to the most accurate prediction within this framework.

\section{Acknowledgements}
I thank Georg Weiglein and Silja Thewes for the collaboration on this project and Alison Fowler for the first steps of the gNWA in her PhD thesis. Many thanks also go to Aoife Bharucha for useful discussions.
I am thankful for the funding from the Studienstiftung des deutschen Volkes.



\bibliographystyle{elsarticle-num}
\bibliography{ICHEP}






\end{document}